\documentclass[iop] {emulateapj}
\shorttitle{Galactic Distribution Superbubbles}

\shortauthors{Higdon and Lingenfelter}

\begin{document}

\title{THE GALACTIC SPATIAL DISTRIBUTION OF OB ASSOCIATIONS AND THEIR SURROUNDING SUPERNOVA-GENERATED SUPERBUBBLES}

\author{J. C. Higdon}
\affil{W. M. Keck Science Center, Claremont Colleges, Claremont,
CA 91711-5916} \email{jhigdon@kecksci.claremont.edu}
\and
\author{R. E. Lingenfelter}
\affil{Center for Astrophysics and Space Sciences, University of
California San Diego, La Jolla, CA 92093}
\email{rlingenfelter@ucsd.edu}

\begin{abstract}

Core collapse supernovae of massive ($\geq$ 8 M$_{\odot}$)  stars are formed primarily in OB associations and help blow giant superbubbles, where their collective shocks accelerate most of the Galactic cosmic rays. The spatial distribution of these stars is thus important to our understanding of the propagation of the observed cosmic rays. In order to better model the Galactic cosmic-ray distribution and  propagation, we construct a three-dimensional spatial model of the massive-star distribution based primarily on the  emission of the H II envelopes surrounding the giant superbubbles which are maintained by the ionizing radiation of the embedded O stars. The Galactic longitudinal distribution of the 205 $\mu$m N II radiation emitted by these H II envelopes is used to infer the spatial distribution  of superbubbles. We find that the Galactic  superbubble distribution is dominated by the contribution of massive star clusters residing in the spiral arms.

\end{abstract}

\keywords{cosmic rays: propagation --- elementary particles
--- gamma rays: theory --- interstellar medium -- nucleosynthesis
--- supernovae}

\section{INTRODUCTION}

Supernovae (SNe) created by the  core collapse of massive stars  have long been recognized (e.g.,  Baade \& Zwicky 1934; Zwicky 1939; Hayakawa 1956;  Ginzburg \& Syrovatskii 1964; Zatsepin \& Sokolskaya 2007) as the major source of Galactic cosmic-ray acceleration.  However, in the eighty years since then, subsequent observations have demonstrated that collapse supernovae are quite complex. Presently, the principal categories of core-collapse supernovae (CCSNe), classified by their optical spectra,  are: II-P  (plateau), II-L (linear),  IIn (narrow emission lines), IIb (transitional), Ib, and Ic  (e.g., Filippenko 1997). The minimum stellar mass necessary to produce a core collapse supernova has been found to be 8 $\pm$ 1 M$_{\odot}$ from the discoveries of red supergiant progenitors  in ${\it Hubble \ Space  \ Telescope}$ (HST) pre-explosion images of SN II-P (e.g., Smartt 2009).  At solar metallicity an 8 M$_{\odot}$ star corresponds to the spectral classification of B3 V (Drilling \& Landolt 2000). We estimate the main-sequence age of such a 8 M$_{\odot}$ star $\approx$ 3.5$\times$10$^{7}$ yr via  interpolation from a tabulation of mass-dependent main-sequence ages (Massey \& Meyer 2001).

The most massive CCSNe progenitor most likely is associated with SN Ic. The progenitors of  SN Ic  are very massive stars which have been stripped of both their hydrogen and helium envelopes (e.g., Filippenko 1997). However stellar progenitors have not been identified on SN Ic pre-explosion images (e.g., Smartt 2009).  SN Ic  models require the core collapse of either rotating single stars initially $\sim$ 40--120 M$_{\odot}$ (Georgy et al.\ 2009),  which have lost their stellar envelopes prior to collapse via stellar winds (e.g., Gaskell et al.\ 1986), or of initially less massive ($\sim$ 25 M$_{\odot}$) primaries (Dessert et al.\ 2012) in close binary systems which have lost their stellar envelopes prior to collapse to smaller companion stars through Roche lobe overflow or common envelope evolution (e.g., Pazy\'{n}ski 1967; Podsiadlowski et al.\  1992). At solar metallicity zero-aged 25 and 120 M$_{\odot}$ stars correspond to  spectral classifications of O8 V and O3 V; these are short lived with main-sequence ages  of 6.4$\times$10$^{6}$  and 2.6$\times$10$^{6}$ yr respectively (Massey 2003).  

These  SN II-P and SN Ic constitute 48 $\pm$ 6\% and 15 $\pm$ 4\%  respectively of CCSNe in spiral galaxies (Smith et al.\ 2011).

These CCSNe progenitors are not distributed uniformly in spiral galaxies. It has been known for close to a century that massive stars cluster (Kapteyn 1914).  This is not surprising since the majority of stars are born in stellar clusters embedded in molecular clouds (McKee \& Williams 1997; Lada \& Lada 2003; Portegies Zwart et al.\ 2010). Moreover, the vast majority (70--80\%) of all O stars reside still  in the stellar clusters in which they were born  (Gies 1987; Mason et al.\ 1998; Neugent \& Massey 2011). The giant  molecular clouds creating the massive CCSNe progenitor stars  usually produce $\sim$5 coeval episodes of star formation separated by $\sim$3.5$\times$10$^{6}$ yr (Blaauw 1964; McKee \& Williams 1997). Together these sub-clusters constitute what is called an ${\it OB}$ ${\it association}$ (OBA) (Blaauw 1964).

Analyses of ${\it Hipparcos}$ proper motion measurements have found  that the upper limit on  internal OBA velocity dispersions  is $\sim$3 km s$^{-1}$ (de Zeeuw et
al.\ 1999).  This modest velocity dispersion is the fundamental reason why CCSNe occur primarily in superbubbles (SB) since  even an 8 M$_{\odot}$ SN II-P stellar progenitor does not travel too far ($\sim$100 pc in 3.5$\times$10$^{7}$ yr) from its birth site before it dies as a CCSNe. Therefore the core-collapse deaths of (120-8 M$_{\odot}$) progenitor stars occur within $\sim$10 to 100 pc of their birth sites. Each stellar death releases $\sim$ 10$^{51}$ ergs of energy in the surrounding interstellar medium with a mean pre-supernova wind contribution of less than 10\% (Leitherer at al.\ 1999). 
Consequently, such correlated supernovae create giant ($>$ 150 pc) cavities of hot, tenuous plasma, or superbubbles, rather than many smaller isolated remnant bubbles (e.g., McCray \& Snow 1979; Tomisaka \& Ikeuchi 1986; McCray \& Kafatos 1987; Mac Low \& McCray 1988; Tenorio-Tagle \& Bodenheimer 1988; Korpi et al.\ 1999; de Avillez \& Breitschwerdt 2005;  Stil et al.\ 2009).   In such superbubbles cosmic rays are accelerated by CCSNe shock waves expanding through tenuous plasmas  heavily enriched by supernova ejecta, synthesized by previous generations of CCSNe  (Higdon et al.\ 1998;  Higdon \& Lingenfelter 2003, 2005; Ogliore et al.\ 2009; Rausch et al.\ 2009).

A concrete example of a  superbubble is the Orion-Eridanus superbubble. Heiles et al.\ (1999) term this superbubble the ``Rosetta Stone of Superbubbles." Its overall size is $\sim$250 pc,  it is filled with a x-ray ($\sim$2$\times$10$^{6}$ K) emitting plasma, and it is powered by stellar winds and CCSNe of massive stars created in earlier star-formation episodes in the Orion OB1 association   (e.g., Heiles et al.\ 1999; Heiles 2001). This large  plasma volume  is surrounded by a denser partial, expanding ($\sim$15 km s$^{-1}$) shell which is  ionized by the O stars of the Orion OB1 association (Reynolds \& Ogden 1979).  This H II  partial shell is  in turn  enveloped  by an expanding H I shell of mass $\sim$3$\times$10$^{5}$ M$_{\odot}$ (Brown et al.\ 1995).

In the present investigation we create a three-dimensional Galactic distribution of OB associations and their surrounding superbubbles.

\subsection{Outline}

Although the actual determination of the Galactic  spatial distribution of superbubbles is straightforward, all the details involved  make the derivation lengthly and complicated.

In Section 2 we investigate the relationship between CCSNe occurrences and sites of massive star formation in other spiral galaxies. Of particular interest is the study by Anderson et al.\ (2012) of the correlation of SN II occurrences and extragalactic star-formation sites as traced by near ultraviolet continuum emissions. Such  emissions are sensitive  indicators of  the presence  of moderate-mass (8 -- 20 M$_{\odot}$)  SN II stellar progenitors.

In Section 3 we discuss the spectral signatures specific to superbubbles. We find that most appropriate SB signature  turns out to be  N II far infrared lines which have been resolved by the Far Infrared Absolute Spectrophotometer (${\it FIRAS}$) aboard the Cosmic Background Explorer ${\it COBE}$  (Fixsen et al.\ 1999).  We expect that such far infrared lines are radiated by $\sim$10$^{2}$ pc sized  (1 -- 10 cm$^{-3}$) photoionized envelopes surrounding the hot, tenuous superbubbles. These H II envelopes absorb the majority of ionizing radiation emitted by the embedded OB associations (e.g., McKee \& Williams 1997).

In Section 4 we first consider a simple three-dimensional N II emissivity model and simulate the ${\it FIRAS}$ 205 $\mu$m intensities, as a function of Galactic  longitude,  assuming that the  spatial distribution of Galactic superbubbles follows that of an axisymmetric disk population. This modeled ${\it FIRAS}$ 205 $\mu$m intensity, displayed in Figure 1, poorly reproduces the measured   ${\it FIRAS}$ 205 $\mu$m intensity.  More complicated source models are clearly required.

Therefore in section 5 we present three-dimensional N II emissivity distributions based on models consisting of four spiral arms (e.g.,  Vall\'{e}e 2008).  We find that such  spiral-arm distributions can reproduce the measured  (Fixsen et al.\  1999) ${\it FIRAS}$ 205 $\mu$m intensity once the contribution of two nearby ($\sim$ kpc) H II regions are included.

In Section 6 we discuss the fact that the SB spatial distribution  does not scale linearly with the N II 205 $\mu$m emissivity distribution due to the presence of a large-scale radial gradient in the ratio of interstellar nitrogen to hydrogen abundances. Here we present the spatial distribution of  solely the massive star population.  This distribution represents the best measure of the Galactic spatial distribution of superbubbles and their their embedded OB associations.

In Section 7 we interpret the very bright 205 $\mu$m  ${\it FIRAS}$ features at Galactic longitudes, around  80$^{\circ}$  and 265$^{\circ}$, not as a  spiral-arm tangents, but
as nearby H II sources. We model the strong 205 $\mu$m  ${\it FIRAS}$ feature at Galactic longitude,  $\approx$ 80$^{\circ}$ as  a  H II envelope of a superbubble, excited by a  rich embedded OB association, Cyg OB2.   We model the other  205 $\mu$m  ${\it FIRAS}$ feature at Galactic longitudes,  $\approx$ 265$^{\circ}$, as a component of the Vela Nebula, a nearby ($\leq$ 0.35 kpc) H II region,  primarily excited  by a single close O star, $\eta$-Puppis.

Finally, in Section 8 we summarize the results of this investigation.

\section{THE RELATIONSHIP BETWEEN CCSNE OCCURRENCES AND STAR FORMATION SITES}

The majority ($\sim$70\%) of CCSNe are expected to be mid and early B stars, since massive stars follow a Salpeter initial mass function (e.g.,  Bastian et al.\ 2010) and the minimum  CCSNe progenitor mass  is  8 M$_{\odot}$ (e.g., Smartt 2009).  Yet Galactic OBA membership is invariably determined by bright  O stars
(Gies 1987; Mason et al.\ 1998; Ma\'{i}z-Apell\'{a}niz et al.\ 2004). As noted by  Ma\'{i}z-Apell\'{a}niz et al.\  ``the quality of the data regarding membership to a cluster or an OBA is extremely varied and for stars located at farther distances information is usually poorer."  Cluster membership of the early and mid B stars, which constitute the majority of the CCSNe stellar progenitors are seldom investigated.  In this section we discuss the evidence for stellar clustering of lower-mass, B-star ($<$ 20 M$_{\odot}$) CCSNe progenitors.

James \& Anderson (2006) investigated the correlation of extragalactic CCSNe to star formation sites using  H$\alpha$ CCD images of the parent galaxies. In order to moderate contributions of faint star-formation sites they employed ${\it all}$ the pixels of an H$\alpha$ image in their analysis. They sorted all the pixels of a digitized image in order of increasing pixel counts starting with the most negative sky pixels (created by background subtractions) and ending with the pixel corresponding to the brightest star-formation region. The initial negative pixel values are set to zero. Then  a cumulative distribution is created from the ranked sequence and is normalized to the total H$\alpha$ flux of the image.  Thus a  NCR (their acronym for the normalized cumulative rank pixel function) value of 0 corresponds to background sky values, and NCR of 1 corresponds to brightest pixel  in an image.

If the CCSNe progenitor was a member of the massive-star population which creates the tracer (here H$\alpha$) emission, it is equally likely to come from any part of the NCR distribution (James \& Anderson 2006). Thus over a long period of time the large number of pixels constituting the 0.0 to 0.1 NCR segment  contribute as much to total tracer emission as the few bright pixels comprising the 0.9 to 1.0 segment.  In such a case the NCR distribution will be uniform and have a mean value of 0.5 (James \& Anderson 2006).

Using the above  NCR pixel statistic, Anderson et al.\ (2012) investigated the relationship between CCSNe and star-formation sites, traced by H$\alpha$ emission, for a large sample of 260 (163 SN Type II and 96 Type Ib/c)  CCSNe. Their effective pixel size corresponds typically to a dimension of 130 pc.  In view of their sample size they applied the NCR statistic to the six sub-type classifications (II-P, II-L, IIn, IIP, Ib, \&  Ic) applicable to CCSNe. Their results are listed in Table 1, where n refers to the number of SNe in the sample.   Using the Kolmogorov-Smirnov (KS) test they found that three NCR distributions,  SN IIb and SN II-L, as well as SN Ic,  agree well with  a massive-star population as traced by H$\alpha$ emission.  Yet, the KS test demonstrates that two Type II sub-types, SN II-P (the most common CCSNe) and SN IIn, as well as SN Ib, are not consistent with a massive-star population as traced by H$\alpha$ line emission.  However, as we shall see, H$\alpha$ emission is not the sole tracer of massive stars.

\begin{center}
Table 1. Anderson et al.\ (2012) \\
\begin{tabular}{|l|c|c|} \hline
\multicolumn{3}{|c|}{Correlations of CCSNe with  H$\alpha$ Tracer}\\  \hline
SNe Class     &   n    & $<$NCR$>_{H\alpha}$   \\   \hline
Ib    &   39    & 0.32  $\pm$ 0.040   \\   \hline
Ic    &   52     & 0.46 $\pm$  0.040  \\    \hline
II-L  &   13 & 0.37 $\pm$  0.100   \\  \hline
IIb   &    13 & 0.40 $\pm$ 0.100    \\   \hline
II-P  &   58 & 0.26 $\pm$  0.039   \\  \hline
IIn   &   19 & 0.21 $\pm$ 0.065   \\ \hline
\end{tabular}
\end{center}

Anderson et al.\ also investigated further the correlation of CCSNe and star formation sites,  but now used near ultraviolet  (NUV 1750--2750 \AA)  images taken by the ${\it Galaxy}$ ${\it  Evolution}$  ${\it Explorer}$ ${\it  (GALEX)}$  (Martin et al.\ 2005).   NUV continuum emission is particularly well suited to identify $\geq$ 10 Myr old stellar clusters, which radiate negligible Lyman continuum (Lyc)  emission.  Such UV emission  provides a ${\it direct}$ measure of star formation on timescales $\sim$10$^{8}$ yr (Martin et al.\ 2005).  Note that a $\sim$10$^{8}$ yr lifetime corresponds to 5 M$_{\odot}$ B5 V star (Massey \& Meyer 2001). Using NUV emissions they found for the SN II-P (50 SNe) $<$NCR$>_{NUV}$ = 0.50 $\pm$ 0.045  and SN IIn (18 SNe) $<$NCR$>_{NUV}$  =  0.38 $\pm$ 0.065.  The KS test shows  that these two NCR distributions are  consistent with that expected for flat distributions, and, consequently, these two SN II sub-types track well stellar clusters containing older (10--100 Myr), but still massive (5--20 M$_{\odot}$) stars.

Thus analyses of sites of extragalactic CCSNe demonstrate that the great majority of these CCSNe occur in or close to ($\leq$ 150 pc) of stellar clusters of massive stars all with the potential to generate superbubbles.

\section{MODELING OF THE GALACTIC SUPERBUBBLE SPATIAL DISTRIBUTION}

We are interested in radiation signatures specific to superbubbles. The  clearest indicator of superbubbles is soft x-ray emission from  large ($\geq$ 10$^{2}$ pc) cavities filled with hot ($\sim$10$^{6}$ K) plasmas. Yet photoelectric absorption by intervening interstellar gas hinders the detection of such soft x-ray emissions  beyond the solar vicinity. In their seminal study of Galactic OB associations  McKee \& Williams (1997) noted that  moderate-density ($\sim$ 3 cm$^{-3}$) photoionized H II envelopes surrounding hot, tenuous superbubbles cores  absorb the majority of the ionizing radiation emitted by embedded association O stars.   Therefore the determination of the spatial distribution of these H II envelopes provides an accurate measure of the distribution of OBA,  the primary Galactic CCSNe sites.

Previously Anantharamaiah (1986) proposed such H II envelopes  as the primary Galactic source of the  diffuse low frequency ($\sim$325 MHz)  radio recombination lines (RRL). Subsequent analyses of a larger RRL data set by Roshi \&  Anantharamaiah (2001)  strengthened the interpretation of moderate-density H II envelopes as the origin of  low-frequency RRL:  first,  RRL emissions exhibit large fluctuations as a functions of longitude, which is  expected as lines of sight intercept discrete  H II envelopes of differing properties,  second, the path lengths through the individual clumps range from 20 to 200 pc, and,  third,  the mean electron densities, n$_{e}$, are $\sim$1-10 cm$^{-3}$.   Consequently, for the reasons presented here we will employ observations of H II envelopes as tracers of Galactic superbubbles.

In their investigation of OB associations McKee and Williams (1997) employed far infrared N II line emissions together with thermal radio continuum radiation to determine the Galactic  Lyc emission of OB associations. We will use far infrared N II line emissions to determine the spatial distribution of H II envelopes of superbubbles. Recently, Steiman-Cameron et al.\  (2010) have employed such N II line emissions to determine the Galactic spatial distribution of photoionized gas. Our study differs from this investigation in several important respects, which will be discussed later.

The rationale for using  far infrared N II line emissions to trace H II emission is the following. Since the ionization potentials of H I and N I are 13.6 eV and 14.5 eV, H II and N II regions are essentially coincident. Thus  we use the N II far infrared lines  at 122 $\mu$m \& 205 $\mu$m as observed by the Far Infrared Absolute Spectrophotometer (${\it FIRAS}$) aboard the Cosmic Background Explorer ${\it COBE}$  (Fixsen et al.\  1999) to infer the spatial distribution of these moderate-density H II envelopes. Note that the ratio of 122 $\mu$m line flux to 205 $\mu$m line flux is density dependent and it rises with increasing electron density, n$_{e}$. In the limit of very low densities, where all the nitrogen ions are in the ground state,  the ratio of the 122 $\mu$m line flux to 205 $\mu$m line flux is 0.7;  the  ratio rises from 1 to 1.6  for n$_{e}$ between 10 and 35 cm$^{-3}$ for uniform-density H II regions  (Wright et al.\ 1991).  Yet at n$_{e}$ of 10$^{2}$ cm$^{-3}$ the ratio increases to 3  and at even higher densities it saturates at 10 (Rubin 1985). However  Fixsen et al.\ found that ${\it FIRAS}$ instrumental noise precluded a detailed derivation of 122 $\mu$m line flux but the overall Galactic disk mean ratio of 122 $\mu$m line flux to 205 $\mu$m line flux is 1.1 $\pm$ 0.1. Thus the ${\it FIRAS}$ N II flux ratio infers that typical electron densities similar to the values expected (e.g., McKee \& Williams 1997; Roshi \&  Anantharamaiah 2001)  for H II envelopes.

Where are such superbubbles expected to be  located in the Galaxy? In their investigation McKee \& Williams (1997) made the simplifying assumption that the surface density of OB associations could be approximated by an axisymmetric Galactic disk population, S($\rho$) $\propto$  e$^{-\rho/H_{\rho}}$, where S represents the population's surface density and 
$\rho$   a source planar distance from the Galactic center.   They determined that  H$_{\rho}$ = 3.34$^{+ 0.71}_{-1.24}$ kpc when they used
an investigation of luminous Galactic  H II regions (Smith, Biermann, \& Mezger 1978), which found that ${\it all}$ bright disk H II regions are restricted to $\rho_{min} \leq  \rho \leq \rho_{max}$ where $\rho_{min} = $ 0.39R$_{s}$, $\rho_{max}$ = 1.3R$_{s}$, and R$_{s}$ is the distance of the Sun from the Galactic center.  We now model the longitudinal ${\it FIRAS}$ 205 $\mu$ intensity distribution corresponding to an axisymmetric truncated exponential disk.

\section{CALCULATION OF THE SPATIAL DISTRIBUTION OF THE  N II EMISSIVITY FOR A TRUNCATED AXISYMMETRIC DISK SOURCE MODEL}

Such a source distribution is best represented in a Galactocentric Cartesian coordinate system,  where the y-axis is aligned along the Sun-Galactic Center line,  and the x-axis, which passes through the Galactic Center (GC), is parallel to the galactic longitude l = 90$^{\circ}$.  There a  Galactocentric cylindrical ($\rho$, $\theta$, z) coordinate system is parameterized as $\rho$ = $\surd$(x$^{2}$+y$^{2}$), $\theta$ = tan$^{-1}$(y/x), and z is measured normal to the plane.  We employ the above S($\rho$) to represent the disk N II 205 $\mu$m emissivity and then determine the corresponding longitudinal distribution of the N II 205 $\mu$m intensity.  The N II intensity must be calculated in Earth–-centered coordinates (r, l, b), where r is distance, l is Galactic longitude, and b is Galactic latitude, which are related to the previously defined Galactocentric cylindrical coordinates ($\rho$, $\theta$, z),
\begin{eqnarray}
\rho = \sqrt{r^2cos^2b+ R^2_s -2R_scosbcosl}, \\
 \theta = tan^{-1}( \frac{R_s - rcosbcosl}{rcosbsinl}\ ), \  z = rsinb. \nonumber
\end{eqnarray}
The N II 205 $\mu$m line  emissivity, $\epsilon^{D}$,  in units of erg s$^{-1}$ cm$^{-3}$, is calculated from S($\rho$) in the following way,
\begin{eqnarray}
\epsilon^{D}(\rho) = L_{N II 205 \mu m}  S(\rho) P_{z}(z)/A_{D},
\end{eqnarray}
where the Galactocentric planar distance $\rho$(r, l, b), is a  function of Earth-centered (r, l, b), L$_{N  II  205 \mu m}$ is the total Galactic 205 $\mu$m luminosity in erg s$^{-1}$,   P$_{z}$(z) represents the height dependence of the H II SB envelopes, and A$_{D}$ represents the source-weighted Galactic-disk area.  Once P$_{z}$(z) is defined, the disk emissivity can be used to calculate a line energy flux, F,
\begin{eqnarray}
F \! = \! \int^{l_{max}}_{l_{min}} \! \! \int^{b_{max}}_{b_{min}} \! \! \int^{r_{max}}_{0} \frac{ \! \! \epsilon^{D}(\rho(r, l, b))}{4 \pi r^2}\ r^2 d r dl cosbdb,
\end{eqnarray}
in units of  erg cm$^{-2}$ s$^{-1}$.

Using ${\it COBE}$,  Bennett et al.\ (1994) performed a  far-infrared spectral line survey from over 99\% of the sky with a 7$^{\circ}$ beam, resolving the Galactic longitudinal distributions of 122 and 205 $\mu$m N II  lines,  see their Fig.\ 2d and 2e. Subsequently, Fixsen et al.\  (1999) analyzed further the ${\it FIRAS}$ data and produced a factor of 2 improvement in the signal to noise ratio of the line emissions as well as a more complete sky map of the N II 205 $\mu$m line emissions.  Consequently, we employ the later analyses of Fixsen et al.\ (1999) in comparisons with our intensity models.  In their Fig.\ 5e Fixsen et al.\  (1999) plot the mean intensity in units of  W m$^{-2}$ sr$^{-1}$ of the N II fine-structure ground-state transition at 205 $\mu$m averaged over Galactic longitudes (centered on the Galactic plane) of $\Delta$l = 5$^{\circ}$ and $\Delta$b = 1$^{\circ}$.  As noted by Fixsen et al.\  (1999) the 7$^{\circ}$ ${\it FIRAS}$  beam does not resolve  the latitudinal distributions of the far infrared lines, consequently we compare the ${\it FIRAS}$ N II 205 $\mu$m line measurements with our models of the mean line intensity, averaged over $\Delta$b = 1$^{\circ}$ and $\Delta$l = 5$^{\circ}$ in the following way,
\begin{eqnarray}
 \frac{d^2F}{dldb}\  \! \! = \! \! 
\frac{1}{\Delta b} \int^{+3.5^{\circ}}_{-3.5^{\circ}} \! \! \!  \int^{r_{max}}_{0} \frac{\epsilon^{D}(\rho(r, l, b))}{4\pi r^2}\ r^2 d r cosbdb,
\end{eqnarray}
in units of erg cm$^{-2}$ s$^{-1}$ sr$^{-1}$.

Before we calculate our modeled disk intensity,  we must address how we represent P$_{z}$(z). Here we assume a Gaussian distribution as a function of z,
 \begin{eqnarray}
                                                       P_{z}(z) = \frac{e^{-0.5z^2/\sigma_z^2}}{\sqrt{2\pi} \sigma_z}\ .
\end{eqnarray}
As noted above,  ${\it FIRAS}$ observations did not resolve the latitudinal extents of the far infrared line fluxes, yet the low-frequency RRL observations (Roshi \&  Anantharamaiah 2001)  found that in the inner Galaxy the low-frequency RRL intensities as a function of Galactic latitude, b, possess full-width half-maxima (FWHM) of $\sim$1.8$^{o}$. In order to reproduce such FWHM transverse the Galactic plane our modeling of the diffuse Galactic N II 205 $\mu$m line emissions required $\sigma_{z} \approx$ 0.15 kpc, significantly greater than $\sim$ the 0.045 kpc, scale height of OB stars (e.g., Reed 2000) or the 0.035 kpc, scale height of molecular clouds (e.g.,  Stark \& Lee 2005). But our $\sigma_{z}$ of 0.15 kpc is consistent with the scale height expected (McKee \& Williams 1997) for H II envelopes.

 Our Figure 1 shows a modeled ${\it FIRAS}$ 205 $\mu$m mean intensity distribution corresponding to an axisymmetric truncated exponential disk with a radial scale, H$_{\rho}$,  of 2.5 kpc. Note that the radial gradient of the N II emissivity distribution is steepened (e.g., McKee \& Williams 1997) by the presence of a radial gradient in the interstellar nitrogen to hydrogen abundance ratio.  Thus a massive star distribution with a radial scale, H$_{\rho}$ of 3.4  kpc, creates a steeper N II emissivity distribution with an H$_{\rho}$ of 2.5 kpc due to the existence of a large-scale Galactic gradient in N/H,  e$^{-\rho/H^{N/H}_{\rho}}$ with a H$^{N/H}_{\rho}$ of 10 kpc (Daflon \& Cunha 2004).  We normalized this modeled ${\it FIRAS}$ 205 $\mu$m intensity to the measured line intensity at l = 30$^{\circ}$ which required a total Galactic N II 205 $\mu$m luminosity of 1.85$\times$10$^{40}$ erg s$^{-1}$. Inspection of this Figure shows that such a modeled intensity, created by an axisymmetric disk distribution,  decreasing monotonically in $\rho$, falls off too slowly with Galactic longitude $>|30^{\circ}|$ and fails to reproduce the   observed ${\it FIRAS}$ longitudinal intensities.  More complicated emissivity models are necessary.

\section{CALCULATION OF THE SPATIAL DISTRIBUTION OF THE N II EMISSIVITY FOR SPIRAL-ARM SOURCE MODELS}

Based on optical observations Morgan et al.\ (1952) were the first to discern that H II regions traced out spiral arms in the Milky Way Galaxy. In a later groundbreaking  study, using both H$\alpha$ and high-frequency RRL observations of H II regions, Georgelin \& Georgelin (1976) found that the Galactic H II distribution delineated a spiral structure comparable to the large-scale spiral configurations observed in other galaxies and  the majority (80\%) of the luminous Galactic H II regions fell along  two symmetrical pairs of arms with pitch angles of 12$^{\circ}$. Subsequent studies of spiral galaxies  have demonstrated that the principal spiral-arm tracers are massive star formation sites defined by  massive stars, photoionized gas and giant molecular clouds (e.g., Baade 1963; Boulanger et al.\ 1981; Rumstay \& Kaufman 1983; Russeil 2003; Hou et al.\ 2009; Efremov 2011).

Since that first pioneering investigation by Morgan et al.\ (1952) Galactic spiral arms have been traced primarily by H II regions (e.g., Georgelin \& Georgelin 1976; Downes et al.\ 1980; Caswell \& Haynes 1987; Watson et al.\ 2003; Russeil 2004, 2007;  Paladini et al.\ 2004;  Hou et al.\ 2009) as well as by giant molecular clouds (e.g., Cohen et al.\ 1986; Dame et al.\ 1986; Grabelsky et al.\ 1988; Digel et al.\ 1990; May et al.\ 1997; Heyer et al. 2001; Dobbs et al.\ 2006; Stark \& Lee 2006; Nakanishi \&  Sofue 2006; Hou et al.\ 2009).  However, other tracers such as H I (e.g., Englmaier \& Gerhard 1999; Nakanishi \& Sofue 2003; Levine et al.\ 2006) as well as infrared emissions (e.g., Hayakawa et al.\ 1981; Bloemen et al.\ 1990; Drimmel \& Spergel 2001; Benjamin et al.\ 2005) also have been used to determine some spiral-arm properties, particularly, the locations of spiral-arm tangents.

In a detailed study evaluating and critically weighing  the past evidence of Galactic spiral structure, Vall\'{e}e (2008) has compiled key spiral-arm parameters and their uncertainties and, consequently, he determined up-to-date locations of four logarithmic spiral arms.  We employed Vall\'{e}e's spiral-arm parameters in our models of N II line ${\it FIRAS}$ intensities. We have expanded Vall\'{e}e's model by the inclusion of the fall offs of the spatial density of H II envelopes parallel to the spiral arms and normal to them.  In the present section we now model the spatial distribution of the superbubble spiral-arm population.

Vall\'{e}e's spiral-arm model can be described in the Galactocentric cylindrical coordinate system, presented earlier, by
\begin{eqnarray}
\rho (\theta ) = {\rho_{min }}{e^{k(\theta  - {\theta _{\circ}})}} \\
\rho_{min} \le \rho  \le \rho_{max}, \nonumber
\end{eqnarray}
where k = tanp, and p is the pitch angle.  From Vall\'{e}e's Fig.\ 1 we determined the following arm parameters for the starting points of the arms: $\rho_{min}$ of 2.9 kpc,  $\theta_{\circ}$ of 70$^{\circ}$  (Norma-Cygnus arm), 160$^{\circ}$ (Perseus arm), 250$^{\circ}$ (Sagittarius-Carina arm), and 340$^{\circ}$ and (Scutum-Crux arm). For the spiral-arm pitch angles, p,  we employed a value of 13.5$^{\circ}$, the unweighted mean value from  Vall\'{e}e's Table 1 for three spiral arms (Norma-Cygnus, Sagittarius-Carina, and Perseus).  In order to reproduce\footnote{Using a pitch angle, p, of 13.5$^{\circ}$ for the Scutum-Crux spiral arm produced arm tangents at l = 26$^{\circ}$ and l = 315$^{\circ}$; changing p to 15.6$^{\ circ}$ placed arm tangents at l = 27$^{\circ}$ and l = 311$^{\circ}$.} the observed ${\it FIRAS}$ 205 $\mu$m line flux at  l = 310$^{\circ}$ we used a somewhat greater pitch angle, p, of 15.5$^{\circ}$ for the Scutum-Crux arm in order to place its tangent at this longitude.  We employ Vall\'{e}e's  $\rho_{max}$ of 35 kpc. Our rendition of Vall\'{e}e’s arm model is shown in our Figure 2. We use R$_{s}$ = 7.6 kpc,  the distance of the Sun from the Galactic center.

In our investigation we employ the above Eq.\ (6) to represent the ${\it medians}$ of the spiral arms and introduce three parameters to construct a three-dimensional spatial distribution for a spiral-arm population. First, we modeled the decrease of the arm population with Galactocentric radius with an exponential distribution
\begin{eqnarray}
                                                       P_{\rho}(\rho) = e^{-\rho/H_{\rho}}.
\end{eqnarray}
McKee \& Williams (1997) noted that such an exponential distribution with H$_{\rho}$ $\approx$ 3.3 kpc reproduced well the axisymmetric Galactic spatial distribution of young stellar populations as well as that of giant molecular clouds.   Second, following Taylor and Cordes (1993) modeling of the spiral-arm component of Galactic thermal electrons, we represent the fall off of arm populations across an arm with distance, $\Delta$, transverse to the arm medians of the modeled arms, as Gaussian distributions,
 \begin{eqnarray}
                                                       P_{\Delta}(\Delta) = e^{-0.5\Delta^{2}/\sigma^{2}_{A}},
\end{eqnarray}
with a fixed dispersion, $\sigma_{A}$.   Third, we employed a Gaussian distribution, Eq.\ 5,  as well as $\sigma_{z}$ of 0.15 kpc, to represent the fall off of H II envelopes with height, z, transverse the Galactic plane.  Now we derive realistic values of H$_{\rho}$ and $\sigma_{A}$ from model comparisons with the Galactic N II 205 $\mu$m longitudinal intensities of Fixsen et al.\  (1999).

For every pair of chosen arm parameters, H$_{\rho}$ and $\sigma_{A}$, the corresponding planar arm distribution is calculated using the following approach. In the Galactocentric coordinate system, defined earlier, coordinate pairs ($\rho_{k}, \theta_{k}$) are calculated along an arm median from Eq.\ 1. Then at each $\rho_{k}$, Eq.\ 7 is used to calculate relative arm source density, P$_{\rho}$($\rho_{k}$).   At these latter locations, further positions,  ($\rho_{j}, \theta_{j}$), are calculated at fixed distances, $\Delta_{j}$, projected normal to each arm. At these positions, Eq.\  8  is used to calculate relative arm density, P$_{\Delta}$($\Delta_{j}$), at each $\Delta_{j}$ ranging non-uniformly from 0.01 kpc to 5 kpc. At the start ($\rho_{k}$ = $\rho_{min}$) and the termination ($\rho_{k}$  = $\rho_{max}$) of an arm, where our knowledge of arm properties is particularly fragmentary, each P$_{\Delta}$($\Delta_{i}$) is projected in a circular arc, and a function of $\Delta_{j}$, as is illustrated in our Fig.\ 2 for$\Delta_{j}$ = 0.5 kpc. Thus the relative arm source densities, P$_{\rho}$($\Delta_{k}$)$\times$P$_{\Delta}$($\Delta_{j}$), are calculated for a sample of 10$^{5}$ locations in the vicinities of the arms.

The fractional contribution of each spiral arm to the total N II Galactic luminosity
( in erg s$^{-1}$)  is found by integrating the product, P$_{\rho}$($\Delta_{k}$)$\times$P$_{\Delta}$($\Delta_{j}$), over the Galactic-disk area and normalizing the surface integrals to N II luminosity fractions discussed below f$_{X}$, where X represents NC (Norma-Cygnus), SA (Sagittarius-Carina), SC (Scutum-Crux), and P (Perseus).   In order to use such modeled arm distributions,  interpolants, S$_{X}$($\rho, \theta$), were created (for each arm X) to permit the calculation of planar source densities as functions of arbitrary Galactocentric coordinates($\rho, \  \theta$),
\begin{eqnarray*}
S_{X}(\rho, \theta) = f_{X}P_{\rho}(\rho_{k}(\rho, \theta))\times P_{\Delta}(\rho_{j}, \theta_{j})/A_{X},
\end{eqnarray*}
\begin{eqnarray*}
A_{X} = \int^{2\pi}_{0}\int^{35 kpc}_{0} P_{\rho}(\rho_{k}(\rho, \theta))P_{\Delta}(\Delta_{j}(\rho, \theta)) \rho d \rho d  \theta.
\end{eqnarray*}
Note that each S$_{X}$($\rho$, $\theta$) distribution, when integrated over the disk area normalizes to f$_{X}$, and, thus, the sum of these four integrated distributions equals unity.

We now employ  these S$_{X}$($\rho,\  \theta$) to represent the spiral-arm contribution of the N II 205 $\mu$m emissivity and then determine the corresponding longitudinal distribution of the N II 205 $\mu$m intensity.  The N II 205 $\mu$m line  emissivity in each arm, $\epsilon^{X}$  in units of erg s$^{-1}$ cm$^{-3}$, is calculated from S$_{X}$($\rho$,  $\theta$) in the following way,
\begin{eqnarray}
\epsilon^{X}(\rho, \theta, z) \! = \!  L_{N II 205 \mu m}  S_{X}(\rho(r, l, b), \theta(r, l, b)) P_{z}(z),
\end{eqnarray}
where the Galactocentric planar distance, $\rho$(r, l, b),  and the polar angle, $\theta$(r, l, b),  are functions of Earth-centered (r, l, b), L$_{N II205 \mu m}$ is the total Galactic 205 $\mu$m luminosity in erg s$^{-1}$.  Using Eq.\ 5 to model P$_{z}$(z),  these arm emissivities can be used to calculate  mean line intensity, averaged over $\Delta$b = 1$^{\circ}$, in the following way,
\begin{eqnarray}
 \frac{d^2F_{X}}{dldb}\  \!  \! = \!
\frac{1}{\Delta b} \int^{+3.5^{\circ}}_{-3.5^{\circ}}  \! \ \! \int^{r_{max}}_{0} \frac{ \! \! \epsilon^{X}(r, \theta, z)}{4 \pi r^2} \! \! \ r^2 d r cosbdb,
\end{eqnarray}
in units of erg cm$^{-2}$ s$^{-1}$ sr$^{-1}$.

In Fig.\ 3 intensities, d$^{2}$F/dldb, summed over all four arms, calculated from Eq.\ 10, are shown for a  total Galactic  N II 205 $\mu$m luminosity of 1.4$\times$10$^{40}$ erg s$^{-1}$ and nominal spiral arm parameters  of $\sigma_{A}$ = 0.5 kpc and a H$_{\rho}$ = 2.4 kpc for four sets of relative arm luminosities,  f$_{SA}$,  f$_{SC}$,   f$_{NC}$, and  f$_{P}$.   The calculations illustrated in  Fig.\ 3 demonstrate that the equal spiral-arm luminosities cannot reproduce the observed ${\it FIRAS}$ 205 $\mu$m  longitudinal intensity fall off at l $\geq$  35$^{\circ}$ and l $\geq$ 335$^{\circ}$ in the  N II 205 $\mu$m intensities and, consequently, the contributions of the Norma-Cygnus and Sagittarius-Carina arms must be decreased by a factor of $\sim2$ relative to the contributions of the Scutum-Crux, and Perseus arms. In their spiral-arm model of far infrared N II line emissions Steiman-Cameron et al.\  (2010) noted similar behavior.

Such differences in the arm contributions have also  been identified in analyses of other spiral-arm tracers.  Drimmel and Spergel (2001) modeled near infrared and far infrared continuum emissions measured by ${\it DIRBE/COBE}$ and they found that two spiral arms (Scutum-Crux and Perseus) dominate the  infrared spiral-arm signatures and that the contribution of the Sagittarius-Carina arm must be reduced relative to  the contribution of the Scutum-Crux arm. Such a two-armed continuum infrared spiral signature is supported by a recent derivation (Benjamin et al.\ 2005; Benjamin 2008) of  the stellar source distribution from the ${\it Spitzer/GLIMPSE}$ mid-infrared survey of the Galactic plane.  Recently  Efremov (2011) noted that both the  Sagittarius-Carina and  Norma-Cygnus arms are better defined  in H I than the other two arms but young stellar populations exist ``beyond a doubt" in the Sagittarius-Carina arm.
Moreover such arm differences have been identified in other spiral galaxies. For example, the derivation of spatial distributions of H II regions in the two-armed Sc galaxy, NGC 628, by Kennicutt \& Hodge (1976) showed that the mean H II densities differ by 1.8 between the two arms.

In view of this discussion and the our modeled intensities illustrated in Fig.\ 3, in the remainder of this study we employ:  f$_{SA}$ = 0.18, f$_{NC}$ = 0.18, f$_{SC}$ = 0.28, and f$_{P}$ = 0.36.

In Fig.\ 4 intensities, d$^{2}$F/dldb, summed over all four arms, calculated from Eq.\ 10, are shown for a total Galactic N II 205 $\mu$m luminosity of 1.4$\times$10$^{40}$ erg s$^{-1}$ for a fixed  $\sigma_{A}$ =  0.5 kpc and a range of  radial arm parameters H$_{\rho}$ from 1.8 to 3.0 kpc showing that the value of 2.4 kpc gives the best fit.  In Fig.\ 5 intensities, d$^{2}$F/dldb, summed over all four arms, calculated from Eq.\ 10, are illustrated for a fixed total Galactic N II 205 $\mu$m luminosity of 1.4$\times$10$^{40}$ erg s$^{-1}$ for a fixed radial arm parameter, H$_{\rho}$, of 2.4 kpc for a range of arm-width parameters, $\sigma_{A}$, from 0.25 to 0.75 kpc showing that the value of 0.5 kpc gives the best fit.  Also we include in these last two figures  our estimates of the contributions of the two nearby H II regions: a $\sim$6$^{\circ}$-sized Cygnus X diffuse H II region with its center at l = 80$^{\circ}$ \& b = 0$^{\circ}$) as well as from 10$^{\circ}$-sized Vela-complex  H II region with its center at l = 262$^{\circ}$ \& b = -2$^{\circ}$). The properties of these two nearby H II regions are addressed in section 7. 

From  these two figures, we find that H$_{\rho}$ = 2.4 $\pm$ 0.3 kpc, and $\sigma_{A}$ = 0.5 $\pm$ 0.1 kpc reproduce well the diffuse ${\it FIRAS}$   L$_{N II 205 \mu m}$ longitudinal intensity, when the two strong local H II sources are included. These parameters this define our three-dimensional  spiral-arm luminosity model for N II  205 $\mu$ m emission, as shown in Fig.\  6 together with the individual contributions from each of the spiral arms, for  a total spiral arm N II  205 $\mu$m luminosity of $\approx$ (1.4 $\pm$ 0.15)$\times$10$^{40}$ erg s$^{-1}$.

 Using these values and our nominal Galactic luminosity N II 205 $\mu$m of 1.4$\times$10$^{40}$
erg s$^{-1}$ our model results suggest that  total Galactic  205 $\mu$m line luminosity inside the solar circle ($\rho \leq$ R$_{s}$) is 1.2$\times10^{40}$ erg s$^{-1}$. From their analyses of ${\it FIRAS}$ line measurements Bennett et al.\  (1994) estimate,  with a $\sim$ 50\% uncertainty, a somewhat greater total 205 $\mu$m Galactic line luminosity inside the solar circle of 2.6$\times$10$^{40}$ erg s$^{-1}$.

As can be seen in Fig.\ 4 and 5 all our models systematically overproduce the observed N II 205 $\mu$m intensity along the tangent to the Sagittarius spiral arm at 55$^{\circ} \leq$  l $\leq$ 30$^{\circ}$, where it is known (e.g., Georgelin \& Georgelin 1976;  Forbes 1983) that  a 5-kpc portion of the Sagittarius is ``devoid" of luminous H II regions.

From these figures we also see that the introduction of a galactic interarm population is not required. A lack of evidence for interarm superbubble population is perhaps surprising.  Recent analyses of high-resolution H$\alpha$ observations of two nearby spiral galaxies shed some light on this thorny issue.   Employing HST images, Lee, Hwang, \& Lee (2011) cataloged 19,598 H II regions  at a resolution of 0.1" (5 pc) in  M51, a system of two interacting galaxies, NGC 5194, a two-armed, grand-design spiral, and a small barred lenticular galaxy NGC 5195.  In NGC 5194 they determined the numbers of H II regions in the arms, nucleus, and the interarm region, respectively 12245, 4422, and 2636. They found that the relatively less luminous interarm H II regions contribute only 8\% of total Galactic H$\alpha$ H II luminosity of NGC 5194.    Recently, Azimlu, Marciniak, \& Barmby (2011) cataloged 3961 H II in the neighboring Sb galaxy, M31, with a resolution of 1" (3.8 pc). They determined that, although $\sim$40\% of the H II regions occurred in the interarm region, the much less luminous interarm H II regions contributed only 15\% of the total Galactic H$\alpha$ luminosity.  Based on such H$\alpha$ observations we suggest that a modest ($\sim$10\%) interarm SB contribution is difficult to discern due to uncertainties in the spiral-arm models as well as in the contribution of nearby H II regions.

Recently  Steiman-Cameron et al.\  (2010) published a spiral-arm source model  to characterize ${\it COBE/FIRAS}$ far infrared emissions for the C II 158 $\mu$m  line as well as for the N II 205 $\mu$m line. Our approach differs from their study  in several important respects. We focus our  investigation solely on 205 $\mu$m radiation from H II interstellar phases, assuming an H II envelope superbubble origin,  and, consequently, we used  as  ancillary data low-frequency RRL observations from Roshi \&  Anantharamaiah (2001). This led to our choice of $\sigma_{z}$ of 150 pc for the Gaussian dispersion for H II envelopes transverse the Galactic plane.  Steiman-Cameron et al.\  did not employ the  H II envelope superbubble origin for N II gas. Instead they used   a $\sigma_{z}$ of 70 pc that is appropriate to
the C II 158 $\mu$m latitude distribution which has been resolved in a balloon experiment (Nakagawa et al.\ 1998). A significant component of the C II 158  $\mu$m line emission may be emitted  (e.g., Bennett et al.\ 1994; Roshi \&  Anantharamaiah 2001) by dense photodissociation regions of molecular clouds which possess a relatively small scale height transverse the Galactic plane (e.g. Stark \& Lee 2003).  Thus we chose a greater value for $\sigma_{z}$.
Further, they convolved their modeled intensities over the ${\it FIRAS}$ beam while we employed the published (Fixsen et al.\  1999) area-averaged intensities. We also represented the geometry of the spiral-arm terminators differently. 

Nonetheless the  ${\it COBE/FIRAS}$ 205  $\mu$m data is robust. Although there exist significant differences in approach, the fundamental results of both investigations are the same:  first, four-armed spiral models with differing relative arm luminosities fit the ${\it FIRAS}$ 205  $\mu$m  longitudinal distributions, and, second, an interarm source contribution is not required.

\section{CALCULATION OF THE SPATIAL DISTRIBUTION OF LYMAN CONTINUUM EMISSION}

Based on this model for the N II 205 $\mu$m emission,  we now calculate the spiral arm density distribution of the Lyc emission from  the O stars which produced it. 
As discussed above the Galactic Lyc does not scale simply with N II 205 $\mu$m emissivity distribution due to the presence  (e.g., Shaver et al\ 1983) of a gradient, as a function of Galactocentric radius, in the interstellar abundance of nitrogen.  Subsequently,  Daflon \& Cunha (2004) analyzed  the emissions of OB stars located at  4.7 kpc $\leq \rho \leq$ 13.2 kpc and derived a  nitrogen abundance gradient of -0.042 dex per kpc or an equivalent H$^{N/H}_{\rho}$ of 10 kpc. Thus from the  spiral-arm N II emissivity distributions
 H$_{\rho}$ = 2.4 $\pm$ 0.3 kpc that  reproduced the ${\it FIRAS}$ 205 $\mu$m line intensities, we expect that 1/H$^{NII}_{\rho}$ $\approx$ 1/H$^{N/H}_{\rho}$ + 1/H$^{LYC}_{\rho}$, and, consequently, the actual fall off in the Galactic Lyc emissivity, H$^{LYC}_{\rho}$ = 3.2 $\pm$ 0.5 kpc.
This is in excellent agreement with the axisymmetric source model of McKee and Williams (1997) where H$_{\rho}$ = 3.3 kpc.  From a radial scale of H$^{N/H}_{\rho}$ of 10 kpc for the interstellar N to H ratio and nominal spiral-arm parameters (H$_{\rho}$ = 2.4 kpc  and $\sigma_{A}$ = 0.5 kpc) we find that the Galactic source weighted N/H = 1.78(N/H)$_{\odot}$.

The resulting spiral-arm Lyc luminosity density model for H$_{\rho}$ = 2.4 kpc  and $\sigma_{A}$ = 0.5 kpc, normalized to a total Galactic Lyc luminosity of unity is shown in FIg.\ 7. This figure represents the best measure of the Galactic spatial distribution of superbubbles and their embedded OB associations.

\section{NEARBY SOURCES OF DIFFUSE 205 $\mu$m N II   EMISSIONS}

Early investigations (e.g., Weaver 1970; Jackson et al.\ 1979; V\'{a}zquez et al.\ 2008) of young stars in the solar vicinity ($\leq$ kpc)  reported that the Sun is``located in an arm-like structure" terminated by the Cygnus region at one end at l $\sim$ 80$^{\circ}$ and Vela at the other end at l $\sim$ 264$^{\circ}$; even then such a feature was termed (e.g., Weaver 1970) a ``spur" or ``offshoot" of the Sagittarius arm.  Vall\'{e}e refers to the structure as the ``Orion armlet" or ``Orion bridge"   describing it  as a ``local aggregate of stellar clusters". This local structure is not viewed as a separate component of the global spiral-arm structure (e.g., Georgelin \& Georgelin 1976; Taylor \& Cordes 1993; Vall\'{e}e 2008). Yet  inspection of the observed ${\it FIRAS}$ 205 $\mu$m longitudinal intensity (Fixsen et al.\ 1999), illustrated in our Figs.\ 1, 3, 4, 5, and 6, show that significant emission at l $\sim$ 80$^{\circ}$ and l $\sim$ 265$^{\circ}$ from this  ``Orion Spur", or ``local aggregate of stellar clusters".

Therefore in addition to the contribution of  the large-scale, Galactic spiral arm distribution in this investigation we consider the emission from these two nearby ($\sim$ kpc) H II regions, the Cygnus X region at l $\approx$  80$^{\circ}$ and the Vela Complex at l $\approx$ 265$^{\circ}$.  A comparison of measured (Fixsen et al.\  1999) ${\it FIRAS}$ intensities at N II 205 $\mu$m  (Fig.\ 5e) and at 122 $\mu$m (Fig.\ 5f) show that the ratios of these two line intensities in the vicinities of l $\approx$ 80$^{\circ}$ and l $\approx$ 265$^{\circ}$ agree with the overall Galactic ratio. Consequently, we assume that the photoionized gas which radiates these lines has approximately the same electron density, $\sim$3 cm$^{-3}$, as the  H II  envelopes in the inner Galaxy which radiate the bulk of the spiral-arm 205 $\mu$m N II line emission.

\subsection{Cygnus Region}

We interpret the bright 205 $\mu$m  ${\it FIRAS}$ feature at l $\approx$ 80$^{\circ}$, not as a  spiral-arm tangent, but as a H II envelope excited by a  rich, nearby OB association, Cyg OB2.   The Cygnus-OB2 stellar cluster is a  visually obscured, but nearby, OB association at a distance of 1.45 kpc (Hanson 2003) with its center at l = 80.1$^{\circ}$ and b = +0.9$^{\circ}$ (Humphreys 1978).  In a groundbreaking study employing infrared observations in the J, H, and K bands Kn\"{o}dlseder (2000) estimated that this rich stellar cluster contained as many as 100 O stars  and, clearly, it is  the closest massive Galactic star formation site.  Recently, employing ${\it Chandra}$ x-ray point-source  data as well as complementary optical and near IR photometry,  Wright et al.\ (2010)  found that Cygnus OB2  encompasses  an estimated 75  O stars as well as $\sim$1200 OB stars.  There is  a large spread of stellar ages, 2 -- 10 Myr, in this rich stellar cluster (e.g., Hanson 2003; Drew et al.\ 2008; Wright al.\ 2010; Comer\'{o}n \& Pasquali 2012).  Most recently, from this massive star-formation site  ${\it Fermi}$ has detected  and resolved a large 4$^{\circ}$  ($\sim$ 10$^{2}$ pc) superbubble, illuminated in 0.1 to 10$^{2}$ GeV $\gamma$-rays emitted by freshly-accelerated cosmic rays in the Cygnus Superbubble (Ackermann et al.\ 2012).  This exciting new observation provides direct evidence that cosmic rays are accelerated in superbubbles.

Lozinskaya et al.\ (2002) investigated the kinematics of photoionized gas in Cygnus region via H$\alpha$ measurements. In this study they posed a fundamental question - ``where are the observational manifestations of the action of intense ionizing radiation from Cyg OB2 on the ambient gas?''  Lozinskaya et al.\  propose that Cygnus X is the ``observational manifestation'' of Cyg OB2. Cygnus X is an extended H II region delineated by an irregular ring of optical H$\alpha$ filaments $\approx$ 8$^{\circ}$ in radius (Ikhsanov 1961).   Other investigates have also suggested Cygnus X and Cygnus OB2 are related (e.g., Ikhsanov, 1961; Veron 1965).

We propose that the H II region, Cygnus X, ionized by O stars in the embedded Cyg OB2,  radiates the bulk of the 205 $\mu$m   line  emission resolved
by ${\it FIRAS}$ from l $\approx$ 80$^{\circ}$.  We estimate the corresponding ionizing luminosity, Q$_{0}$, for this association to be $\approx$ 1.1$\times$10$^{51}$ s$^{-1}$, from  75 young O stars, employing a time-averaged, ionized stellar luminosity relation\footnote{Q$_{0}$(M) = 5.5$\times$10$^{42}$M$^{4}$ s$^{-1}$, for stars born in the mass range, 20 $\leq$ M $\leq$ 40 M$_{\odot}$, and Q$_{0}$(M) = 8.2$\times10^{45}$M$^{2}$ s$^{-1}$, 40 $\leq$  M $\leq$ 120 M$_{\odot}$ (McKee \& Williams 1997).} and a initial stellar mass function with a slope  of  -1.35 (Salpeter 1955).  The Cyg OB2 association is very luminous, exceeding by a factor of $\sim$40, the ionizing luminosity, Q$_{0}$, of the familiar Orion OB association of only 2.7$\times$10$^{49}$ s$^{-1}$ (Williams \& McKee 1997).

Lozinskaya et al.\ (2002) pointed out that the absence of intense H$\alpha$ emissions, expected for such a large number of O stars, could be explained by strong visual absorption, but corresponding thermal radio emissions must be observed regardless of visual obscuration. However, they remark that in the Cygnus X region radio sources are detected against a strong diffuse radio  source $\sim$ 6$^{\circ}$ in size centered on the Cyg OB2 association   (e.g., Wendker 1970; Huchtmeier \& Wendker 1977; Lozinskaya et al.\ 2002), In a 2.7 GHz  radio survey an angular resolution of 11$^{'}$  Wendker (1970) measured a total flux density, S$_{\nu}$, of 5390 Jy for the total emission including resolved sources and solely  2260 Jy for the diffuse component of the Cygnus X.

The ionizing luminosity, Q$_{0}$, required to produce such diffuse thermal radio emission can be estimated (e.g., Afflerbach, Churchwell, \& Werner 1997),
 \begin{eqnarray}
Q_{0} \approx \frac{7.6\times10^{48}}{T^{1/2}_{e}}\  ( \frac{S_{\nu}}{Jy}\ ) ( \frac{\nu}{GHz}\ )^{0.1} ( \frac{r}{kpc} )^2
 \end{eqnarray}
in units of s$^{-1}$ where $\nu$  is the radio frequency in GHz, T$_{e}$ represents the electron temperature. Thus to maintain the diffuse Cygnus X 2.7 GHz radio source at a distance, r,  of 1.45 kpc, T$_{e} \approx$ 8000 K, requires Q$_{0}$ of 0.45$\times$10$^{51}$ s$^{-1}$.   To maintain the entire Cygnus radio emission, discrete sources as well as the diffuse background, requires $\sim$10$^{51}$ s$^{-1}$. In fact the embedded Cyg OB2 massive stars are potentially capable  of the maintaining the entire Cygnus thermal radio emission.

We propose that the $\sim$ 6$^{\circ}$ (R $\sim$ 75 pc)  radio Cygnus X is the source of the N II 205 $\mu$  line emission, resolved by ${\it FIRAS}$, in the Cygnus region.  Further support for this conjecture is provided by  Lozinskaya et al.\ (2002). Approximate the 6$^{\circ}$ Cygnus X source as  a sphere of $\sim3^{\circ}$ radius sphere (R $\sim$75 pc). They remark that a typical electron density is $\sim$4.5 cm$^{-3}$ found for an average  line of sight through such a spherical region, 4R/3 $\sim$100 pc,  and an  emission measure $\sim$ 2000 cm$^{-6}$ pc. Note that their estimated mean density is close to the mean value of 3 cm$^{-3}$ derived by McKee \& Williams for H II envelopes.

Thus, following the lead of Lozinskaya et al., we represent a modeled Cygnus X source, illustrated in Figures 4 to 6, as a uniform sphere with a 75 pc radius, whose center is located at l = 80$^{\circ}$ and b = 0$^{\circ}$. We assume that the source is resolved normal to the Galactic plane. To approximate the effect of the 7$^{\circ}$ beam of ${\it FIRAS}$ instrument, we weighted this source expected intensity over a Gaussian profile with a FWHM of 7$^{\circ}$. To reproduce the observed 205 $\mu$m intensity  in Figures 4 to 6 we found that the required N II line flux of our modeled source is 2.1$\times$10$^{-7}$ erg s$^{-1}$ cm$^{-2}$,  which corresponds to a N II 205 $\mu$m luminosity of 2.4$\times$10$^{37}$ erg s$^{-1}$ at a distance of 1.45 kpc.

This powerful single source contributes 1.7$\times$10$^{-3}$ of the Galactic 205 $\mu$m luminosity.  How does this line luminosity compare to the expected Cygnus contribution to the total number of ionizing photons absorbed by photoionized gas in the Galaxy? Based on analyses of thermal radio emissions and a published study (Bennett et al.\ 1994) of ${\it FIRAS}$ N II far infrared line emissions, McKee \& Williams estimate a total Galactic ionization rate, Q$^{G}_{0}$,  of 1.9$\times$10$^{53}$ s$^{-1}$.  However, based on their prescription this value should decrease by a factor, (R$_{s}$/8.5)$^{2}$ $\sim$0.8, due the decrease in the distance of the Sun from the Galactic Center of 8.5 to 7.6 kpc.  However the derivation of Q$^{G}_{0}$ also relies on the value of L$_{N II 205 \mu m}$, the total Galactic 205 $\mu$m luminosity.  In section 5 we derived this luminosity from a detailed model fit to the ${\it FIRAS}$ 205 $\mu$m longitudinal intensities.  Our 205 $\mu$m luminosity estimate of 1.2$\times$10$^{40}$ erg s$^{-1}$ within the solar circle (R$_{s}$ $\sim$ 7.6 kpc) 
 is half the published value of Bennett et al.\ (1994) of 2.4$\times$10$^{40}$ erg s$^{-1}$  yet our smaller estimate is consistent with their value when $\sim$50\%  uncertainties are included.  Due to difficulties in merging disparate radio data sets, which affect analyses of radio observations of H II regions, we assign greater weight to analyses of far infrared emission lines.  Consequently, we estimate a total Galactic Q$^{G}_{o}$ of 9$\times$10$^{52}$ s$^{-1}$.

If we assume that interstellar gas in the H II shell around Cyg OB2 has solar composition, then its expected fraction of the Galactic Lyc luminosity implied by N II line emissions is $\sim1.78 \ \times 1.7\times10^{-3} = 3\times10^{-3}$.  Thus we estimate that the value Q$_{0}$ required to maintain the Cygnus X N II line emissions to be 2.8$\times10^{50}$ s$^{-1}$. Such an estimated Q$_{0}$ is only 25\% of Q$_{0}$ we calculated from the number of O stars expected to be present as well as the ionizing luminosity required to maintain the total Cygnus X thermal emission. Consequently, we view the nearby Cyg OB2 as the source of N II 205 $\mu$m from the Cygnus region.

\subsection{Gum Nebula -- Vela Region}

We interpret the bright  205 $\mu$m ${\it FIRAS} $ feature at l $\approx$ 265$^{\circ}$, not as a  spiral-arm tangent, but as a H II region excited primarily by a single  close O star, $\eta$-Puppis. The Gum Nebula - Vela Complex contains only three  O stars compared to $\sim$75 O stars residing in
 the rich Cyg OB2  but the Complex is closer by a factor of $\sim$ 5. Also its observed N II 205 $\mu$m line intensity is smaller that $\sim$3 than that of the Cygnus X N II source. The large, nearby Gum Nebula forms an H$\alpha$ ring on the sky with $\sim$18$^{\circ}$ radius with its center at l $\sim$  258$^{\circ}$ and b $\sim$  -2$^{\circ}$ (Gum 1956; Chanot \& Sivan 1983). These H$\alpha$ emissions most likely are powered by stellar UV fluxes from $\eta$ Puppis (O4I), and $\gamma^{2}$ Velorum (WC8+O9I binary) (e.g.,  Gum 1956; Beuermann 1973; Reynolds 1976; Wallerstein, Silk, \& Jenkins 1980; Chanot \& Sivan 1983).

 However, the origin of the Gum nebula itself is controversial. The Gum Nebula has been modeled as: 1) a fossil Str\"{o}mgren sphere, created by the stellar progenitor of the Vela supernova remnant (e.g., Brandt et al.\ 1971; Alexander et al.\ 1971),  2) an old ($\geq$10$^{6}$ yr) supernova remnant, ionized by UV fluxes from  $\eta$ Puppis and $\gamma^{2}$ Velorum (Reynolds 1976), and 3) an ordinary H II region (e.g.,  Gum 1956; Beuermann 1973).  Subsequently, Wallerstein, Silk, \& Jenkins (1980) investigated the properties of interstellar gas in the Gum Nebula using both optical and UV line observations. They found that their data was “generally consistent” with a model of the Gum Nebula as an ordinary H II region, ionized by OB stars,  $\eta$ Puppis, $\gamma^{2}$ Velorum, as well as by the stellar progenitor of the Vela supernova remnant, and the gas is stirred by multiple stellar winds.

Ionized interstellar components with gas densities of  order $\geq$1 cm$^{-3}$, relevant to N II 205 $\mu$m emissivities, can be found in a number of sites in the Gum Nebula. For example, Reynolds (1976) estimated that the H$\alpha$ emission shell is 15 to 30 pc thick and the electron densities $\sim$ 5 -– 2.5 cm$^{-3}$. Recently, Sushch, Hnatyk, and Neronov (2010) investigated the complex interactions of the Vela supernova remnant and the 5$^{\circ}$-sized stellar wind bubble of $\gamma^{2}$ Velorum. They found that the interstellar density around $\gamma^{2}$ Velorum is $\sim$ 12 cm$^{-3}$.

The O-star catalog of Maíz-Apellániz et al. (2004) lists the $\gamma^{2}$ Velorum binary at l = 262.8$^{\circ}$, b = -7.7$^{\circ}$ with spectral types O9I and WC8 and $\eta$ Puppis, a runaway O star, at l = 256$^{\circ}$, b = -4.7$^{\circ}$ with spectral class O4I. The  total stellar Q$_{0}$ is $\sim$1.32$\times$10$^{50}$ s$^{-1}$: $\eta$ Puppis, a very luminous O4I  star, has an expected Q$_{0}$ of 1.05$\times$10$^{50}$ s$^{-1}$ (Vacca et al.\ 1996), the  O9I star of $\gamma^{2}$ Velorum binary has a  Q$_{0}$ of  2.14$\times$10$^{49}$ s$^{-1}$ (Vacca et al.\ 1996), and the WR star of  $\gamma^{2}$ Velorum binary has a  Q$_{0}$ of 6$\times$10$^{48}$ s$^{-1}$ (De Marco et al.\ 2000).   Using radial velocity measurements North et al.\ (2007) determined that $\gamma^{2}$ Velorum binary is close, at a distance, r, of 336$^{+8}_{-7}$ pc.   ${\it Hipparchos}$ parallel measurements provide a distance to  $\eta$ Puppis of 329$^{+120}_{-77}$ pc (van der Hucht et al.\ 1997).  The runaway star, $\eta$-Puppis contributes $\sim80$\% of the total stellar Q$_{0}$ and these two stellar systems are close,  separated only by $\sim$40 pc.

Based on these parameters we assumed a single source of N II line emission,  a uniform sphere with a 30 pc radius, whose center is located at l = 262$^{\circ}$ at a distance of 330 pc.  To approximate the effect of the 7$^{\circ}$ beam of the ${\it  FIRAS}$ instrument we weighted this source intensity over a Gaussian profile with a full-width, half-maximum of 7$^{\circ}$. We found that our model required N II line flux of 7.7$\times$10$^{-8}$ erg s$^{-1}$ cm$^{-2}$, which corresponds to a 205 $\mu$m luminosity of 10$^{36}$ erg s$^{-1}$  at the assumed distance of 0.33 kpc.

The Vela source contributes 10$^{36}$/1.4$\times$10$^{40} \approx$ 7$\times$10$^{-5}$ of the total galactic 205 $\mu$m line luminosity. If we assume that interstellar gas in the vicinity of Vela has solar composition, then its expected fraction of the Galactic Lyc luminosity implied by N II line emissions is $\sim1.78 \ \times 7\times10^{-5} = 1.25\times10^{-4}$. Scaling the Galactic Q$^{G}_{o}$ by 1.25$\times$10$^{-4}$,
we estimate the value Q$_{0}$ required to maintain the Vela N II line emissions to be $\sim 1.1\times10^{49}$ s$^{-1}$.  Since the Vela O-stars actually radiate $\sim$1.3$\times$10$^{50}$ s$^{-1}$, only $\sim$10\% of the stellar ionizing luminosity is required to power the N II 205  $\mu$m emission observed by ${\it FIRAS}$.

\section{SUMMARY}

Core collapse supernovae of OB stars, formed in large OB associations,  blow giant superbubbles, in which their collective shocks accelerate most of the Galactic cosmic rays. With the sites of cosmic ray acceleration now generally recognized, we can now proceed to more sophisticated model tests of how the cosmic rays propagate within the Galaxy.  In order to define the initial conditions for such models of cosmic ray propagation on a Galactic scale, we have constructed a three-dimensional spatial model of the Galactic distribution of superbubbles, using  the measured ${\it FIRAS}$ Galactic longitudinal distribution of the 205 $\mu$m N II luminosity, emitted by H II envelopes surrounding hot  tenuous superbubble cores.  These H II envelopes in turn are excited by the  Lyman continuum radiation of the massive O stars embedded in the superbubble cores.

Recently  Vall\'{e}e (2008) critically evaluated past evidence of Galactic spiral structure and  he determined up-to-date Galactic-plane locations of four spiral arms.  We expanded his  model by the inclusion of the fall offs of the spatial density of H II envelopes both parallel and normal to the positions of spiral arms.  Once suitable choices were made of the parameters governing these density fall offs  we found that  the observed ${\it FIRAS}$ 205 $\mu$m N II longitudinal distribution is  reproduced well by spiral-arm source models once emissions from two nearby H II regions were included. One nearby source is the Cyg OB2 association, one of  the richest Galactic OB associations, located at a distance of 1.45 kpc, with its center at l = 80.1$^{\circ}$ and b = +0.9$^{\circ}$. The other source, at a distance of 0.3 kpc, is ionized by the nearest runaway O star,  $\eta$-Puppis and the O star binary $\gamma^{2}$ Velorum.  Finally, we remark  that, as illustrated by our Figures 4, 5, \& 6  the Galactic  superbubble distribution, as traced by H II envelope emission,  is dominated by the contribution of massive star clusters residing in the spiral arms and two nearby H II regions.

\newpage
\begin{figure}
\begin{center}
\includegraphics[scale=0.5, angle=270]{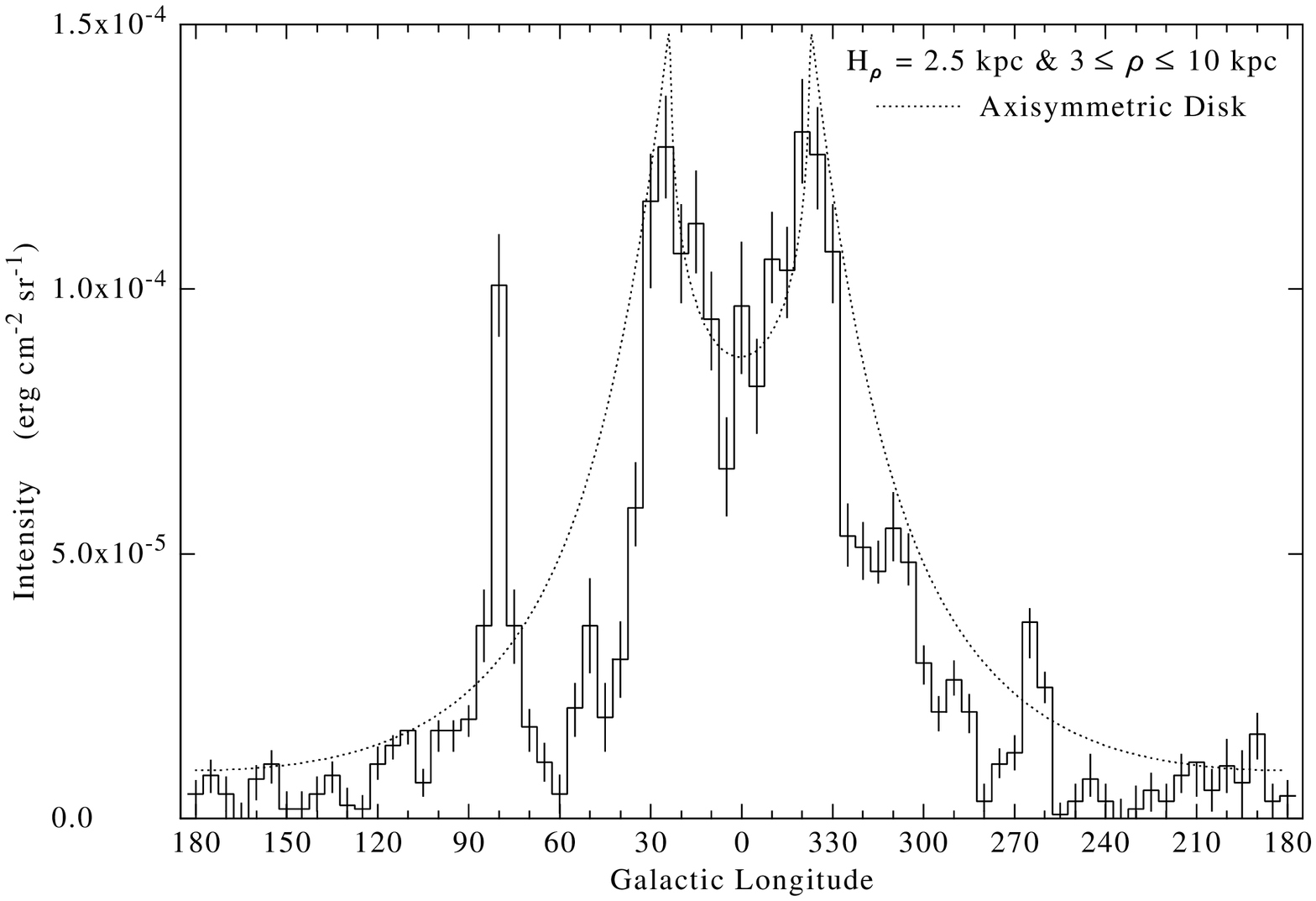}
\end{center}
\caption{Axisymmetric  Disk Source Model. A modeled  N II 205 $\mu$m longitudinal intensity distribution based on a radially truncated axisymmetric disk population with H$_{\rho}$ = 2.5 kpc and R$_{s}$ = 7.6 kpc. A Gaussian distribution transverse the Galactic plane with dispersion, $\sigma_{z}$, of 0.15 kpc is assumed. The modeled intensity distribution is normalized to the observed ${\it FIRAS}$ intensity at l = 30$^{\circ}$ requiring a total Galactic N II 205 $\mu$m luminosity of 1.85$\times10^{40}$ erg s$^{-1}$. Shown as a histogram  are the mean longitudinal ${\it FIRAS}$ 205 $\mu$m intensities,  Fig.\ 5e (Fixsen, Bennett, \& Mather 1999) averaged over Galactic longitudes of $\Delta$l = 5$^{\circ}$ and over Galactic latitudes (centered on the Galactic plane) of $\Delta$b = 1$^{\circ}$.}
\end{figure}

\newpage
\begin{figure}
\begin{center}
\includegraphics[scale=0.75]{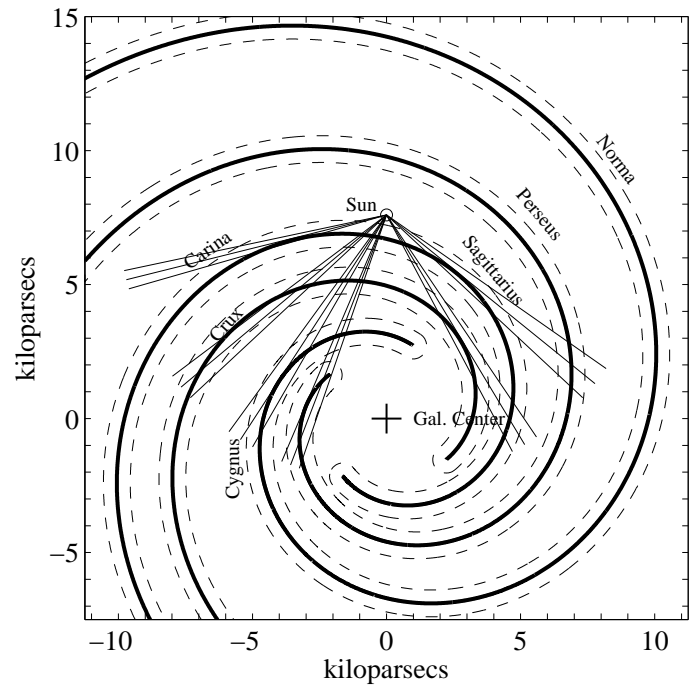}
\end{center}
\caption{Our model of the Galactic distribution of the median of spiral arms based on the prescriptions of Vall\'{le}e (2008). The thick solid curves represent equation (1), for Perseus, Norma-Cygnus, Crux-Scutum, and Sagittarius-Carina arms. Mean longitudes as well as root-mean-square uncertainties of measured spiral-arm tangents from Table 2 of Vall\'{e}e (2008) are represented by thin solid lines. The curves to either side of the arms represent the one $\sigma$ Gaussian 0.5 kpc half-arm widths. }
\end{figure}

\newpage
\begin{figure}
\begin{center}
\includegraphics[scale=0.5, angle=270]{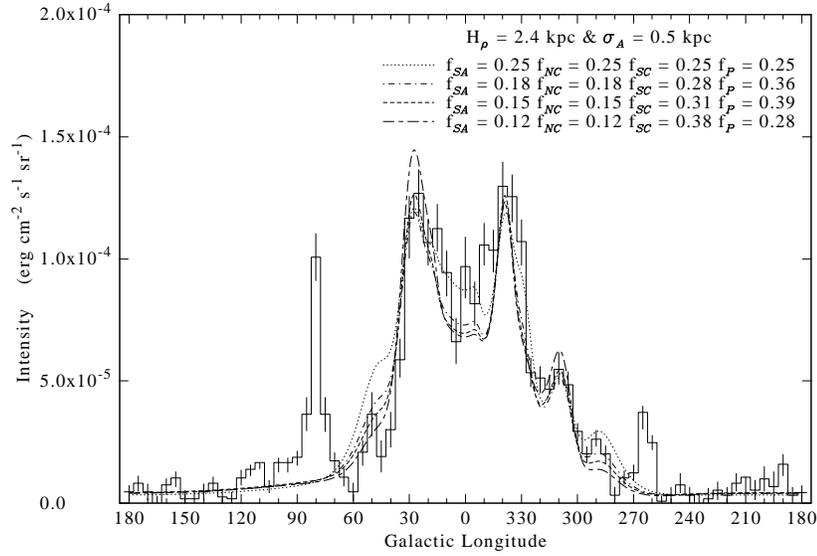}
\end{center}
\caption{Representative 205 $\mu$m N II intensities calculated from Equation (10) for a total Galactic spiral-arm luminosity of  L$_{N \ II \ 205\  \mu m}$ of 1.4$\times$10$^{40}$ erg s$^{-1}$ and nominal  arm parameters of H$_{\rho}$ = 2.4 kpc and $\sigma_{A}$ = 0.5 kpc are displayed. Four different sets of relative arm luminosities,  f$_{SA}$,  f$_{SC}$,   f$_{NC}$, and  f$_{P}$, have been considered and we find that the second set of relative luminosities appears to give the best fit of the four, so as we discuss in the text, we use them hereafter. As in Fig.\ 1 the mean longitudinal ${\it FIRAS}$ 205 $\mu$m intensities, Fig.\ 5e (Fixsen, Bennett, \& Mather 1999) averaged over Galactic longitudes of $\Delta$l = 5$^{\circ}$ are shown as a histogram.}
\end{figure}

\begin{figure}
\begin{center}
\includegraphics[scale=0.5,angle=270]{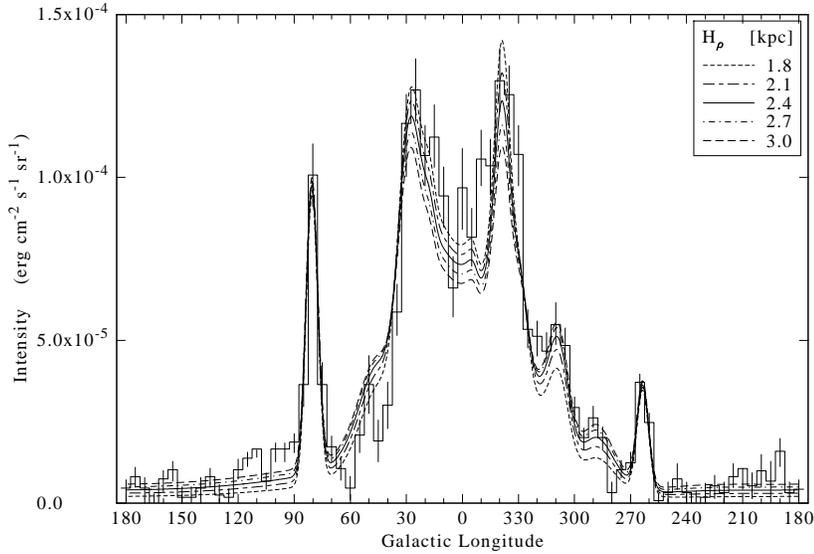}
\end{center}
\caption{Longitudinal Spiral-Arm Distribution as Function of Variations in H$_{\rho}$.  Modeled representative 205 $\mu$m N II intensities calculated for a total spiral-arm luminosity of  L$_{N \ II \ 205\  \mu m}$ of 1.4$\times$10$^{40}$ erg s$^{-1}$ and a fixed arm-width parameter,  $\sigma_{A}$ of 0.5 kpc are plotted for a range of radial parameters, H$_{\rho}$ , between 1.8 to 3.0 kpc, showing the value of about 2.4 appears to give the best fit and we use that hereafter. As in Fig.\ 1 the mean longitudinal ${\it FIRAS}$ 205 $\mu$m intensities, Fig.\ 5e (Fixsen, Bennett, \& Mather 1999) averaged over Galactic longitudes of $\Delta$l = 5$^{\circ}$ are shown as a histogram.   Also included, in addition to the modeled spiral arms, are the contributions of nearby local H II regions, Cygnus X, at l $\approx$ 80$^{\circ}$ and Vela Complex, at l $\approx$ 262$^{\circ}$.}
\end{figure}

\newpage
\begin{figure}
\begin{center}
\includegraphics[scale=0.5,angle=270]{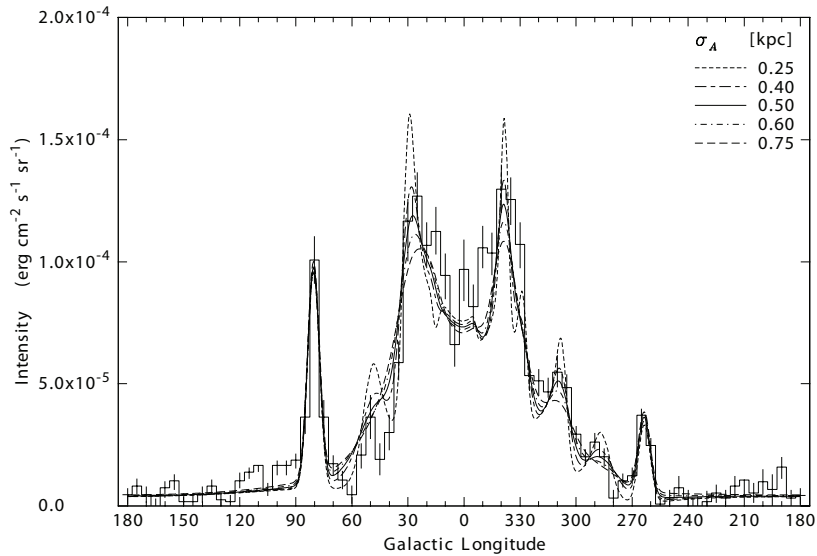}
\end{center}
\caption{Longitudinal Spiral-Arm Distribution as Function of Variations in $\sigma_{A}$. Modeled representative 205 $\mu$m N II intensities calculated for a  L$_{N \ II \ 205\  \mu m}$ of 1.4$\times$10$^{40}$ erg s$^{-1}$   and a fixed radial  parameter, H$_{\rho}$ of 2.4 kpc for a range of arm-width parameters, $\sigma_{A}$, between 0.25 to 0.75 kpc. We find that a Gaussian half-width of 0.5 $\pm$ 0.1 kpc appears to give the best fit and we use that hereafter.  As in Fig.\ 1 the mean longitudinal ${\it FIRAS}$ 205 $\mu$m intensities, Fig.\ 5e (Fixsen, Bennett, \& Mather 1999) averaged over Galactic longitudes of $\Delta$l = 5$^{\circ}$ are shown as a histogram.   Also included, in addition to the modeled spiral arms, are the contributions of nearby local H II regions, Cygnus X, at l $\approx$ 80$^{\circ}$ and Vela Complex, at l $\approx$ 262$^{\circ}$.}
\end{figure}

\begin{figure}
\begin{center}
\includegraphics[scale=0.5,angle=270]{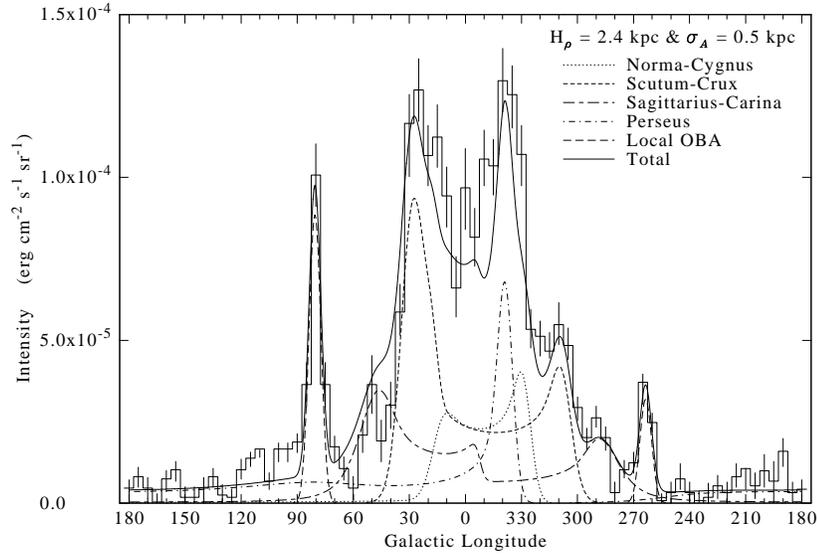}
\end{center}
\caption{Individual Spiral-Arm Contributions to 205 $\mu$m N II Intensities. The solid curve shows the modeled 205  $\mu$m N II intensity distribution for a total spiral-arm luminosity, L$_{N \ II \ 205\  \mu m}$,  of 1.4$\times$10$^{40}$ erg s$^{-1}$  and a fixed arm-width parameter, $\sigma_{A}$, of 0.5 kpc for a radial parameter, H$_{\rho}$, of 2.4 kpc. 
As in Fig.\ 1 the mean longitudinal ${\it FIRAS}$ 205 $\mu$m intensities, Fig.\ 5e (Fixsen, Bennett, \& Mather 1999) averaged over Galactic longitudes of $\Delta$l = 5$^{\circ}$ are shown as a histogram.  Also included, in addition to the modeled spiral arms, are the contributions of nearby local OBA, Cygnus X, at l $\approx$ 80$^{\circ}$ and Vela Complex, at l $\approx$ 262$^{\circ}$.}
\end{figure}

\newpage
\begin{figure}
\begin{center}
\includegraphics[scale=0.5]{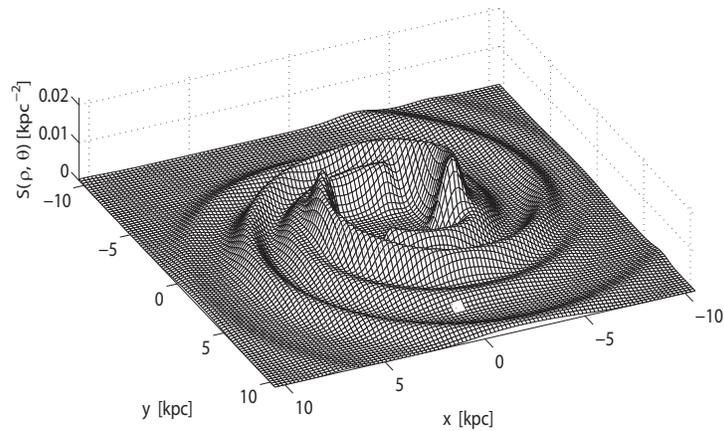}
\end{center}
\caption{A representative planar spiral-arm Lyc source distribution which we also expect to reflect the Galactic
spatial distribution of OB associations and their surrounding superbubbles. A Galactic-centered Cartesian grid with spacings dx = 0.2 kpc and dy = 0.2 kpc was used to create the illustrated arm distribution, corresponding to a nominal H$_{\rho}$ = 2.4 kpc and $\sigma_{A}$ = 0.5 kpc. It was calculated employing the S$_{X}$($\rho$, $\theta$) interpolants, discussed in the text, and the simple transformation x = $\rho$cos$\theta$ and y  =$\rho$sin$\theta$. Note that $\int$S($\rho,\theta$)$\rho$d$\rho$d$\theta$  = 1. For purposes of illustration, the solar position is identified by a white square.}
\end{figure}

\end{document}